\documentclass[twocolumn,showpacs,amsmath,amssymb]{revtex4}

\usepackage{graphicx,rotating}
\usepackage{dcolumn}
\usepackage{bm}
\usepackage{epsfig}
\usepackage{longtable}
\usepackage{color}

\begin{document}

\title{Cascade donor-acceptor organic ferroelectric layers, between graphene sheets, 
for solar cell applications.}

\author{Ma\l{}gorzata~Wierzbowska}\email{wierzbowska@ifpan.edu.pl}
\affiliation{%
Institut of Physics, Polish Academy of Sciences (PAS),
Al. Lotnik\'ow 32/46, 02-668 Warszawa, Poland
}%

\author{Ma\l{}gorzata Wawrzyniak-Adamczewska}
\affiliation{%
Faculty of Physics, A. Mickiewicz University, ul. Umultowska 85, 61-614 Pozna\'n, Poland
}%

\date{\today}

\begin{abstract}
Organic ferroelectric layers sandwiched between the graphene sheets are presented as a model of
the solar cell. The investigated systems display many advantageous properties:
1) the cascade energy-levels alignment, 2) simultaneous donor and acceptor
character depending on the charge-carrier direction, 3) the charge-transfer
excitonic type, 4) the induced polarization of the electrodes, leading to a substantial
work-function change of the anode and cathode - around $\pm$1.5 eV, respectively.     
\end{abstract}

\maketitle

\section{Introduction} 

The power consumption and environmental pollution triggers scientists for 
an intensive search of the efficient photovoltaic materials. 
One of the key factors of the solar-cell efficiency is the supressed 
recombination of the charge carriers. This supression can be achieved via 
introduction of the charge trapping layers \cite{trap},
and/or by the manipulation of the energy-level alignment \cite{cascade}. 
Designing the heterostructure, that exhibits the cascade energy 
levels for the holes and electrons is tricky.
On the other hand, if one applies the multilayers of the same material - which, 
in addition, is ferroelectric - then the Stark effect shifts the energy levels 
of the subsequent layers gradually. Recently published theoretical studies by 
Sobolewski \cite{Sobol}, of the ferroelectric columnar clusters in the context 
of the organic photovoltaics without p-n junctions, concern molecules 
with the dipole moment. The molecular orbitals of the subsequent energy levels,
in that work, are well localized at the corresponding molecular rings in the stack; 
especially for the top and bottom molecules.

Our systems are composed of the flat molecules, named 
1,3,5-tricyano-2,4,6-tricarboxy-benzene. The formula might be written
as C$_6$-3(NCCH$_2$)-3(OCOH). These molecules consist of 
the central aromatic ring. Every second carbon of the ring is terminated 
with the cyano group possessing the dipole moment - standing out of the 
ring planes - and alternated with the carboxy groups, 
which form the intermolecular hydrogen  bonds within the planes.
This is for the purpose of the minimization of the electronic transport within the planes.
The molecule with indexed atoms, as well as the top and side views on 
the single molecular layer are presented in Fig.~1. These networks can be 
stacked one on top of the other in various ways, but for the electronic transport 
the AA-type is favored. Considered structure resemble the 2D covalent metal-organic 
frameworks \cite{Bein-1,Bein-2}, however the intermolecular hydrogen bonds 
are used here instead of the covalent bonds. These molecular layers
are sandwiched between the graphene sheets that act as the electrodes.
The interlayer transport across our stacks is of the $\pi$-$\pi$ type, and it 
strongly depends on the overlap between the $p_z$ orbitals of carbons within the molecular
central part and those of the neighboring molecule, as well as
the distance between the top atoms of the dipole group and 
the bottom hydrogens of the next top molecule.

\begin{figure}
\vspace{5mm}
\centerline{ \includegraphics[scale=0.15]{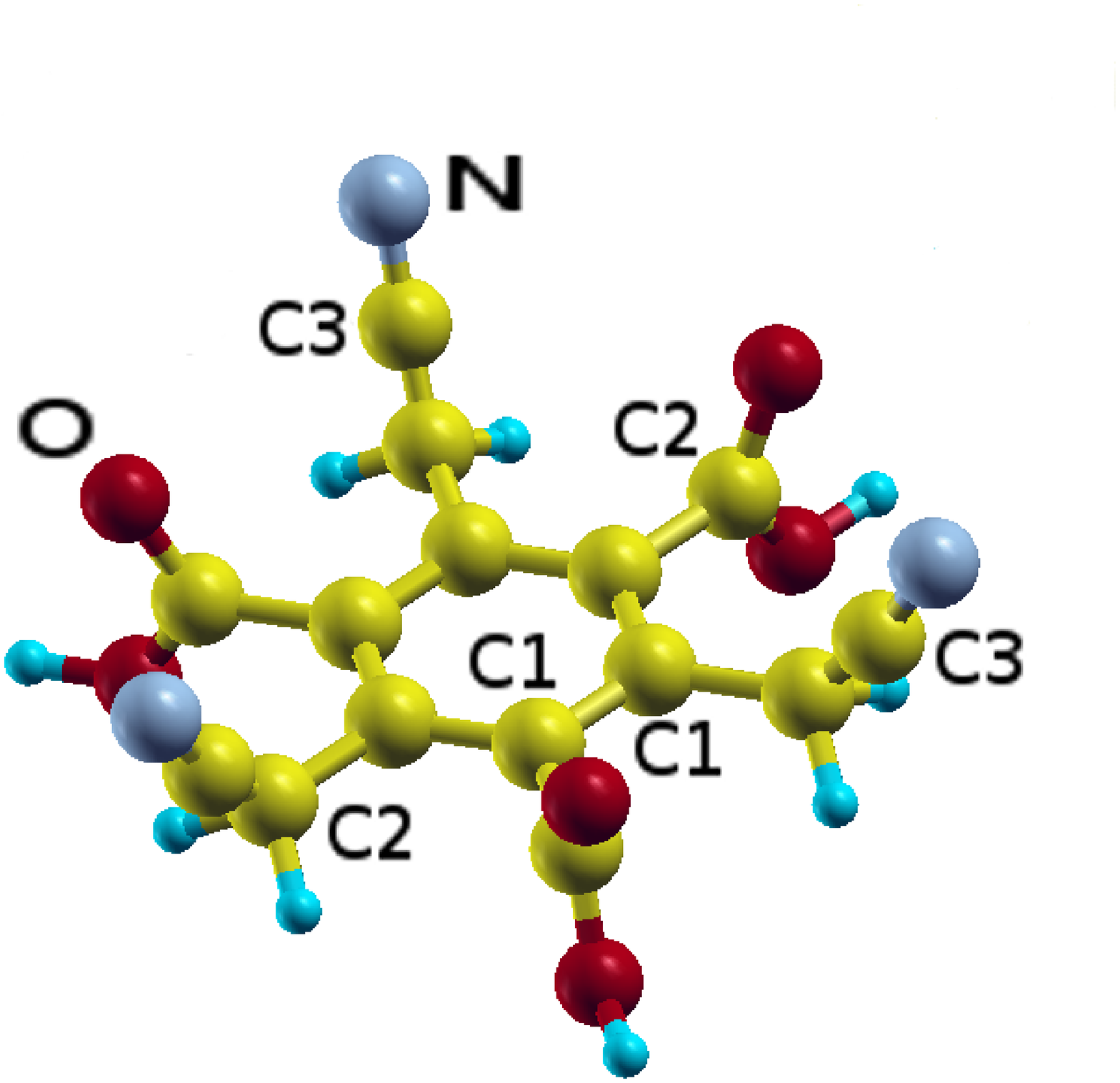}\hspace{1mm}
             \includegraphics[scale=0.2]{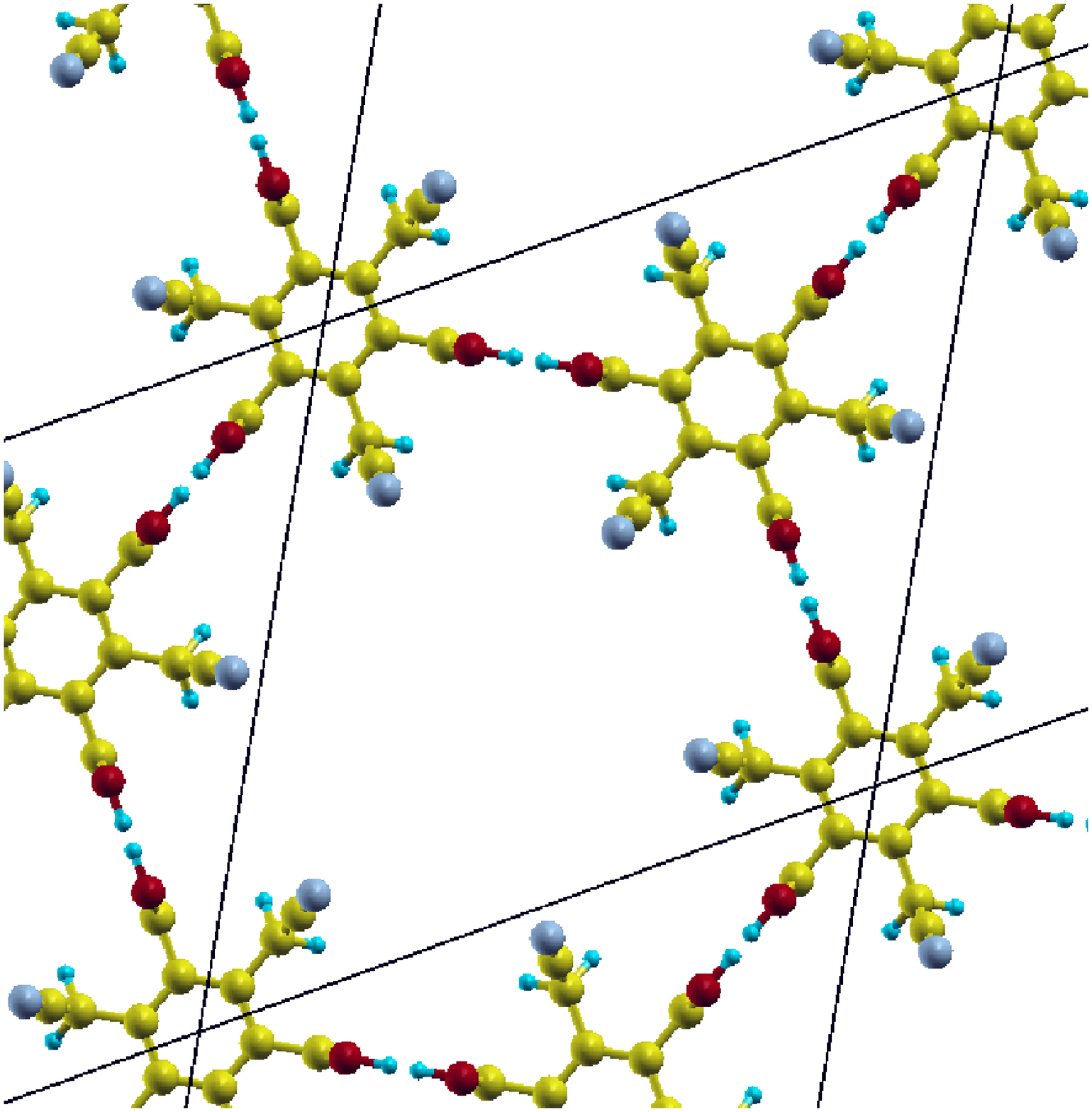}}\vspace{2mm}
\rightline{ \includegraphics[scale=0.2]{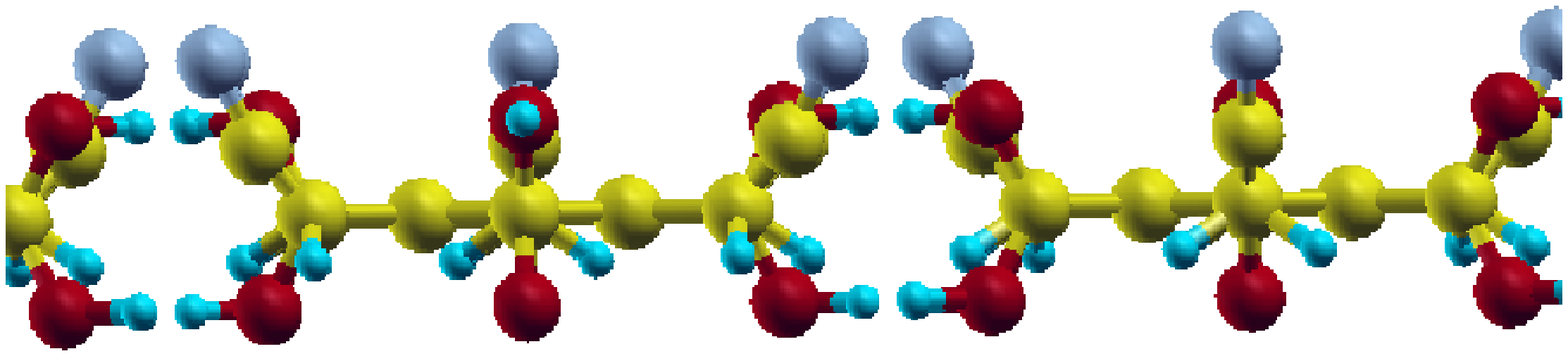} }
\caption{The building molecule with atomic indexes (left panel), top and side
views of the hydrogen bonded hexagonal layer (right panel).}
\label{b1}
\end{figure}

Each ferrolectric monolayer in the stack acts as a donor of the electrons to the
next deeper-laying layer, and simultaneously as an acceptor of the electrons from
neighboring above-laying layer. According to our knowledge, this is the first       
system - except the columns in the work \cite{Sobol} -
where donor and acceptor (D-A) functionality is combined within the same molecule.
Therefore, these layers will operate in a very similar way to the integrated heterojunctions
\cite{Bein-2} or bulk heterojunctions \cite{BHJ,IBHJ},
where two materials of different D-A functionality interpenetrate each other 
in the "bulk" of the optically active region.  

Interestingly, the proximity of the molecular dipole moment 
polarizes the top and bottom electrode in opposite directions.
The induced change of the surface dipole moment influences the work function, 
in such a way that 
it is higher for the anode and lower for the cathode. This is the advantageous feature. 
Similar properties are exhibited by graphene with the deposited ferroelectric layer 
- recent patent \cite{patent1,patent2}, with the documented efficiency of 
the photovoltaic effect about 2$\%$. 
The same idea of the addition of the ferroelectric elements has been applied     
in the 1D quantum-dot solar cells decorated with the molecules possessing the
dipole moment \cite{QCdipole}.

\section{Computational details} 

The density-functional theory calculations were performed using 
the {\sc Quantum ESPRESSO} suite of codes (QE) \cite{qe}. This package is based
on the plane-wave basis set and the pseudopotentials for the core electrons.
The exchange-correlation functional was chosen for the gradient corrected
Perdew-Burk-Erzenhof parametrization \cite{PBE}. The ultrasoft pseudopotentials \cite{USPP}
were used with the energy cutoffs 30 Ry and 300 Ry for the plane-waves and
the density, respectively. 
It has been checked that increasing these values to
40 and 400 Ry, respectively, changes the band energies by less than 0.002 eV
and relative conformation energies of the ferroelectric and antiferroelectric
molecular wire by less than 0.001 eV. 
The Monkhorst-Pack uniform k-mesh in the Brillouin zone has been set to
$4\times 4\times 1$ for the slab with two molecules per the elementary cell,
and $1\times 1\times 10$ for the wire with one molecule per cell
(for two molecules per cell, the corresponding k-mesh was $1\times 1\times 5$),
which was enough due to large supercells.
The vacuum separation between the periodic slabs was around 40 \AA. 

In order to obtain the band structures projected onto the local groups of atoms, 
we employed the wannier90 package \cite{w90}, which interpolates bands using
the maximally-localized Wannier functions \cite{wan,RMP}. The same tool has been used
for the calculations of the dipole moment and polarization. For the Wannier
optimization procedure, we used the k-point mesh of the same accuracy 
as for the self-consistent DFT calculations. The number of Wannier functions 
was the same as the number of electrons in the system. The outer and inner windows
were chosen in such a way, that they were identical and their border energies were
placed within the energy gaps between 
the groups of the composite bands. For metallic
systems, such as molecule between the graphene sheets, the calculation of the Wannier
occupation numbers was necessary. We used the k-point dependent procedure,
being an extension of that used in the van der Waals functionals \cite{VdW}, and
previously tested by us for GaAs doped with Mn \cite{gamnas}.  

The dipole moment can be easily obtained from the maximally-localized-Wannier   
centers positions, $r_n$, using the formula
\begin{equation}
d = \sum_a Z_a R_a - \sum_n r_n
\end{equation}
where $Z_a$ and $R_a$ are the atomic pseudopotential charge and its position, 
and indexes $a$ and $n$ run over the number of atoms and Wannier functions, respectively. 
The above simple Wannier approach is based on the modern theory of
polarization \cite{polar} - which was previously used in the calculations of other 
photovoltaic 2D system \cite{giova}.

\section{Results} 

\begin{figure}
\includegraphics[scale=0.25]{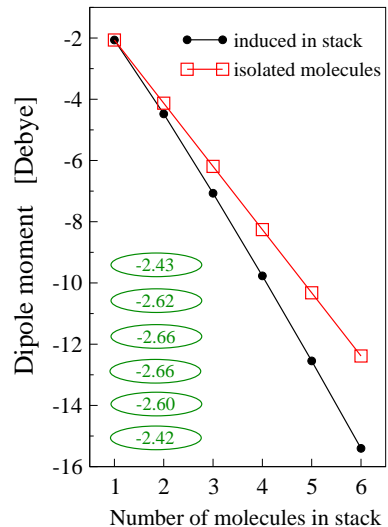}
\caption{Dipole moment induced in the column of molecules depending on their number
in the stack (black circles), compared to the same value for the isolated molecules
(red squares). The dipole moments induced at each molecule, when there are six of them
in the stack, are also given (inside the green ellipses).}
\label{b2}
\end{figure}

\begin{figure*}
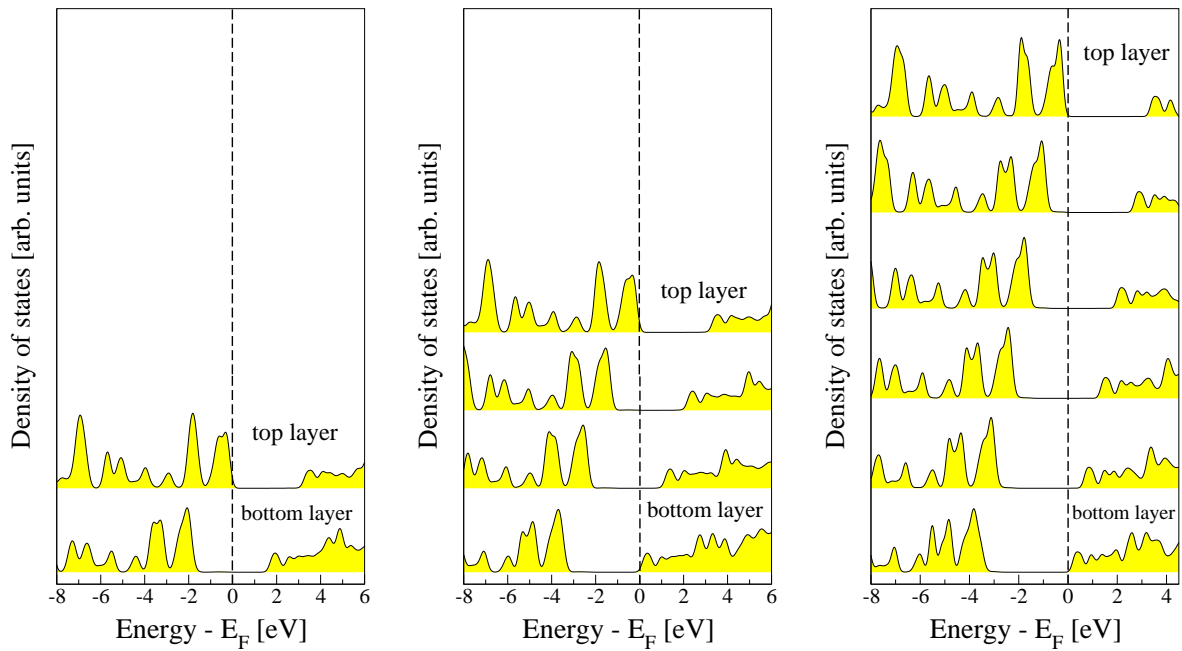

\centerline{ \hspace{-12mm} \includegraphics[scale=0.3]{F3a.eps} \hspace{4mm}
\includegraphics[scale=0.3]{F3b.eps} \hspace{4mm}
\includegraphics[scale=0.3]{F3c.eps} }
\caption{Stark shift of the energy levels visible in
the DOS projected at the layers; for 2-layer, 4-layer and 6-layer cases.}
\label{b3}
\end{figure*}

The geometries of the investigated systems have been optimized, and the intermolecular
distance in the AA-type stack is 5.2 \AA. The distance between the top molecule
and the top graphene is the same as that between the molecules, and on the bottom it
is smaller, of order 4.6 \AA. The lateral intermolecular distance between the molecules
can be described by the OH...O bond, which is 1.75 \AA.

\subsection{Dipole moment.}

Systems presented in Fig.~1 are built of molecules containing two types of groups with
the dipole moment: 1) OCOH and 2) NCCH2. 
The first group is in the plane perpendicular to the plane of the central carbon ring.         
The dipole moment of this group is oriented from O to OH. Due to the fact, that this group
plays the role of chemical connection between the molecules, some molecules have the OH
part above the carbon-ring plane and the neighboring molecules have this group below 
the central part plane. Therefore, in total there is no dipole moment originating from
the intermolecular bonds within the layer, as it is in other hydrogen-bonded molecular
systems \cite{Horiuchi-1,Horiuchi-2}.   
In contrast, the second group plays the role of the polarization source,
with the dipole moment oriented from N down to the central C-ring plane.

The calculated dipole moment of an isolated molecule is -2.06 Debye.
When the molecules are placed in the column, then the proximity of other dipoles
induce larger polarization. The induced dipole moment in the column of molecules
is presented in Fig.~2. The molecules placed in the middle of the finite number stack   
are characterized by the larger induced polarization - about 10$\%$ - than the molecules
at the top and bottom of the column.

We emphasize, the total dipole moment from the multilayer
-- i.e. the monolayers which are in the middle of the stack -- does not cancel as 
in the bulk ferroelectrics. This is because the interactions between the layers
are of the van der Waals type, and molecules do not form the chemical bonds
between the layers. Thus, the electronic polarization at the molecular dipoles 
feels the neighboring layer very weakly. 

\begin{figure}
\includegraphics[scale=0.26]{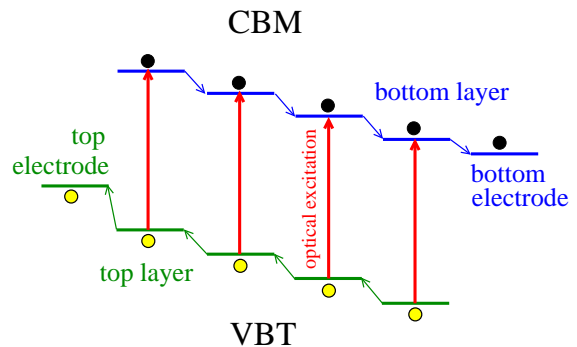}
\caption{Scheme of the photovoltaic mechanism in the cascade energy-level system
with the electrodes. VBT denotes the valence band top and CBM means the conduction
band minimum.}
\label{b4}
\end{figure}

\begin{figure}
\includegraphics[scale=0.3]{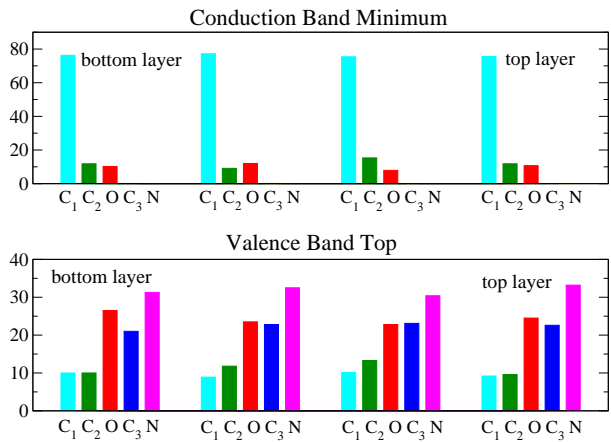}
\caption{The DOS projected at atoms indexed in Fig.~1.
The presented values were taken at maxima of the VBT and CBM peaks,
of the 4-layer system shown in Fig.~3.
The numbers are given in the percentage of the total DOS.}
\label{b5}
\end{figure}

\subsection{Cascade energy-levels alignment and combined donor-acceptor functionality.} 

The energy gap in the band structure of a single molecular layer is about 3.66 eV. 
For the 6-layer thick slab, the single-layer energy gap decreases to around 3.4 eV. 
While the total energy gap of the system closes due to the Stark effect, which shifts 
the top and bottom layer towards higher and lower energies, respectively. 
The intermediate layers shift gradually in the energy, forming a cascade of  
red the hole- and electron-energy levels, 
when moving step by step from the top to the bottom
electrode. In Fig.~3, the density of states (DOS) 
projected onto each layer separately is presented,
for the 2-, 4- and 6-layer stack. The overal energy gap for the 2-layer case is half
of that for the single layer. In the case of 4-layer stack, the total band gap closes.
Increase of the number of molecular layers lowers the step of the energy shift 
of the neighboring layers. The shape of the projected DOS is only slightly changed  
between the layers and as a function of their number. 
In our case, the cascade energy-level alignment is obtained, 
practically without need of band engineering; 
which is the usual case when many materials are used 
in the heterojunctions \cite{cascade}. 
Such cascade is a very desired property in order to avoid the recombination of 
the charge carriers.

Donor or acceptor character of each layer depends on its position with respect to
the next layer - whether the carriers arrive from the anode or cathode - although all
layers have identical atomic structure. Due to our knowledge, this is the first material
which shows the property of donor- and acceptor-type mimetism within the same layer.     
It is advantageous, because one can obtain the functionality of the sophisticated
bulk heterojunctions \cite{BHJ} by using only one 2D material stacked in multilayers.

Fig.~4 schematically shows the operation of the cascade energy-level system with the
electrodes, whose bands energetics follow the sequence of the electron and hole 
energy levels of the optically active material.
Each single optical excitation creates one hole-electron pair 
at one layer. Then, the holes move towards higher energies within a manifold of 
the valence band states and the electrons move down within the conduction band levels.  
This way, the charge carrier transport across 
the optically active layers towards the electrodes takes place.
Its efficiency is additionally factorized by the excitonic character and 
the work function of the molecular layers and the electrodes.

\begin{figure}
\vspace{4.5cm}
\includegraphics[scale=0.28,angle=-90.0]{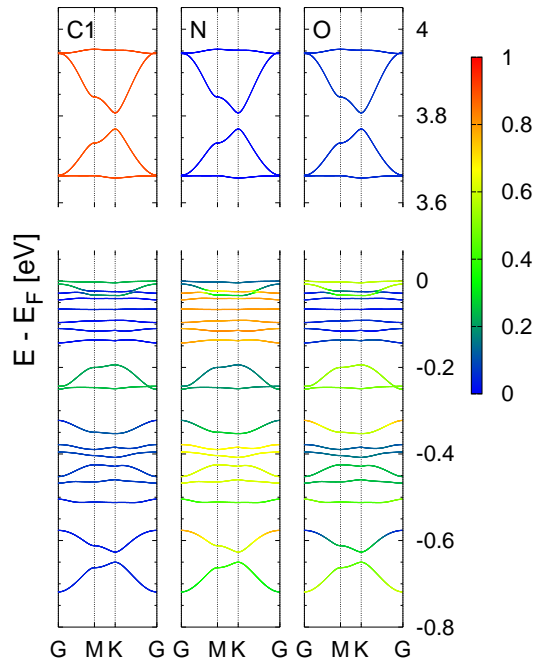}
\caption{Band structure of the single molecular layer projected at the carbons of the
central ring (C1) and nitrogens and oxygens, respectively. The color scale indicates
the factor of the near-atom-centred Wannier functions in the band decomposition.}
\label{b6}
\end{figure}

\subsection{Exciton localization.}

Excitons - the electron-hole pairs - can be described as the difference of 
the charge distribution before and after the absorption of light.
If this difference is spacially localized, we have so-called Frenkel excitons. 
The extended excitations are of the Wannier-Mott type.
The excitons move through the system as a pair before splitting - which might end with
the delayed fluorescence and recombination instead of wanted charge separation.
Moreover, in the molecular or very small quantum dot systems, the excitations 
lead to the singlet state, with the short life-time, or - after irradiative transitions
- to the longer lived triplet state. The excitonic localization character is one of 
the indicators of the photovoltaic efficiency. 

Therefore, we analize the atomic composition of the DOS at the hole and electron peaks 
in the VBT and CBM, respectively. In Fig.~5, we denote the central-ring carbons as C1,
the adjacent carbons as C2, and the farther carbons connected to N as C3. The DOS
projected at these groups of carbons, as well as nitrogens and oxygens, is pictured 
for the peak maximum at the VBT and CBM; 
separately for each molecular plane of the 4-layer system. 
The values are normalized to the total DOS and given in percentage. It is important to
note that the lowest excited electronic states are composed mainly of carbons in 
the central ring. The hole states are delocalized over the dipole groups. The
component from the central part of the molecule to the hole states is only $\sim$10$\%$. 
The above observations are true for all layers in the stack. 
As follows, our electron-hole pairs result from the excitation of the valence electrons
localized around the the anion (groups containing N) into empty states localized
around cation (the central benzene ring). 
Such pairs are called the charge-transfer excitons. A quick estimation for the
Bohr exciton radius in our systems is about 4.7 \AA. It is defined as the distance 
between the center of the molecule and one of the nitrogen atoms.
Therefore, we have Frenkel-type charge-transfer excitonic material.
Interestingly, the charge-transfer excitons are also formed at van der Waals 
interfaces of other materials \cite{CT-vdW}. 

\begin{figure}
\centerline{ \includegraphics[scale=0.24]{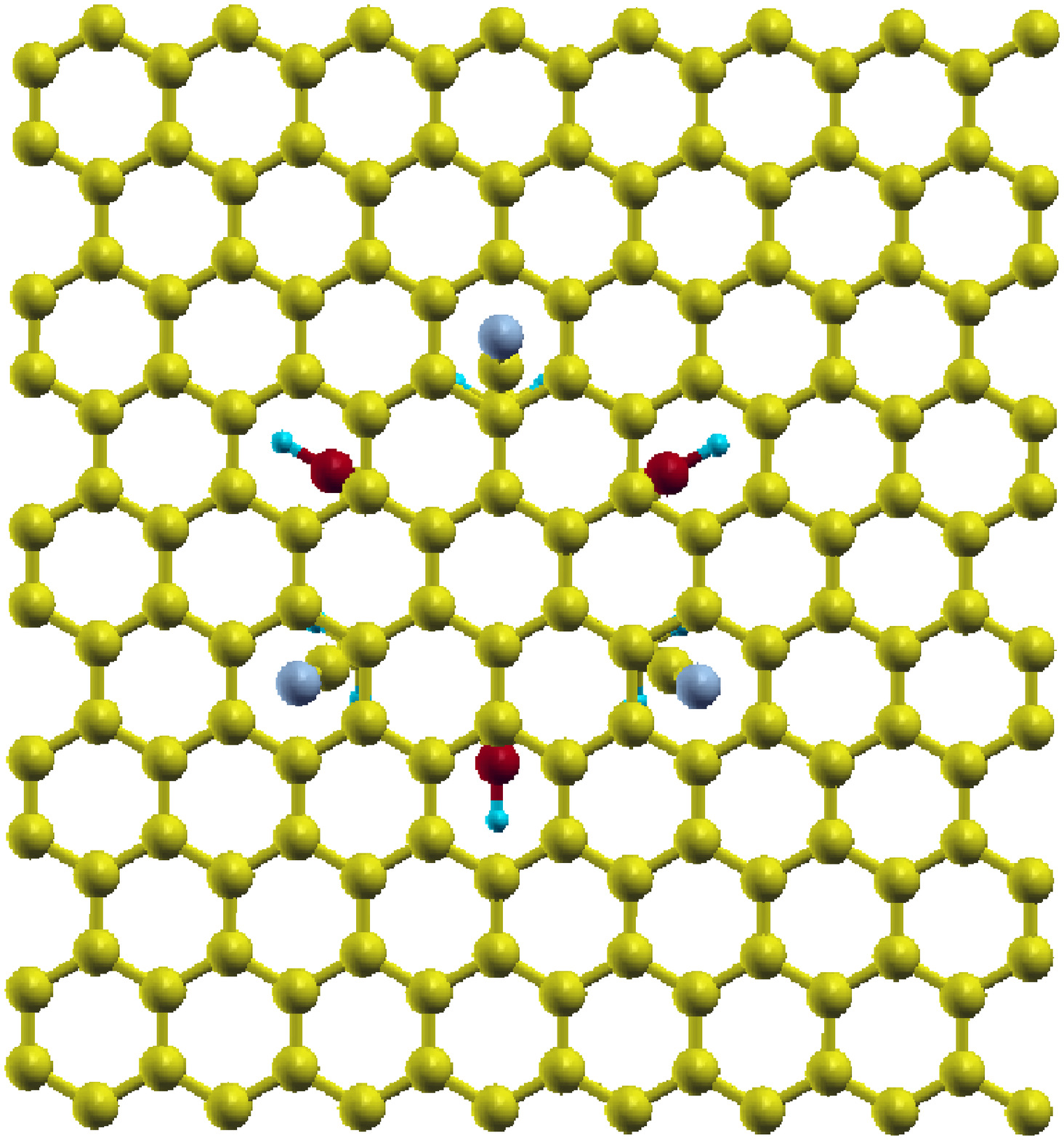}}
\centerline{ \includegraphics[scale=0.24]{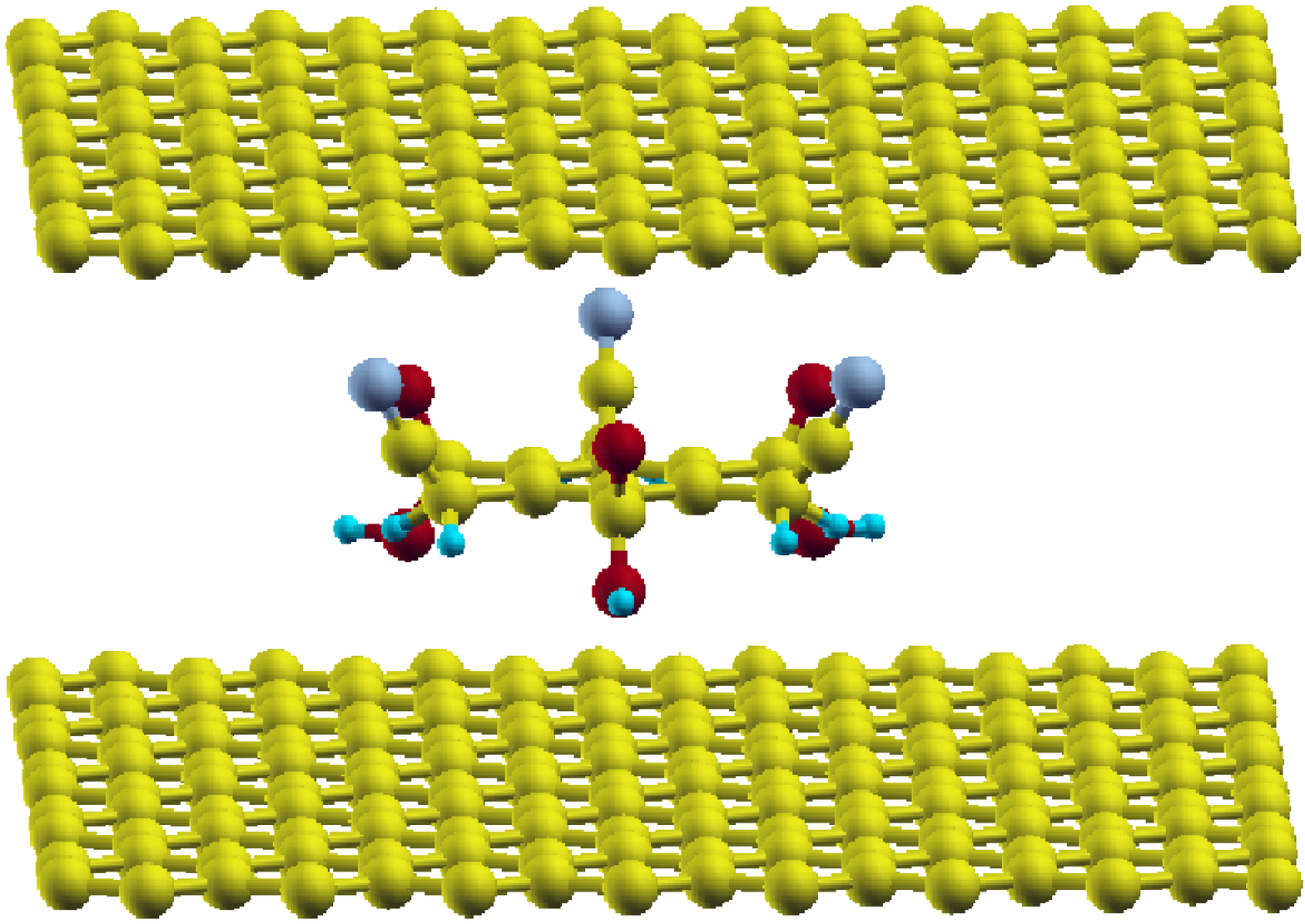} }\vspace{2mm}
\centerline{ \includegraphics[scale=0.3]{F7c.eps} }
\caption{Top and side views at the single molecule between the graphene sheets
and the projected DOS.}
\label{b7}
\end{figure}

For further insight into a localization of electrons and holes, we project
the band structures onto those maximally-localized Wannier functions which
are centred near chosen groups of atoms. These plots are presented in Fig.~6 for
carbons of the central ring, nitrogens and oxygens. The figure confirms our previous
result, that the central-ring atoms are involved in the conduction states and
the dipole and oxygen groups in the valence states. Moreover, we can see that
using the OH...O bonds prohibits 
the electronic and hole transport within the molecular
planes; which is obvious from the flat band structures.   

\subsection{Work function change caused by proximity of ferroelectric layers.} 

The dipole moment of molecules induces the charge polarization within the graphene layers.
In order to estimate the strength of this polarization, we performed the calculations
for the periodic system with large elementary cell, which contains a single molecule
between the graphene layers on top and bottom. The number of carbon atoms in each graphene
part in that cell is 160. The top and side views on this elementary cell are presented
in Fig.~7. The figure contains also the plot of the DOS projected onto the molecule and
the graphene sheets. 
The DOS shift of the top and bottom graphene layers, by about $\pm$0.2 eV 
for each electrode, is caused by the Stark effect and leads to the small charge transfer 
between the leads. 

Using the Wannier-functions technique, described in the "computational details", 
we estimated the induced dipole moment in the top graphene as -2.58 Debye and the
bottom graphene as +2.54 Debye; within the elementary cell with each 
graphene surface around 411 \AA$^2$.      
The above result is promissing for tunning the work function, since the opposite electrodes
polarize differently. It is convenient, for the power of the  photovoltaic conversion,
to chose the anode electrode as a high work function material and the cathode as a low
work function metal.

The experimental value of the work function of the pristine graphene is
 around 4.6 eV \cite{WF-graph}.
It is possible to tune this property with the electric field \cite{WF-E}. The same effect
can be achieved by a decoration of the graphene surface with metals \cite{WF-G-metal}, or  
any 2D system with the molecules with the dipole moment \cite{OH,workmoldipol}. 

The change of the work function $\Delta\Phi$ caused by the induced polarization 
$\Delta P$ is given by the Wigner and Bardeen formula \cite{Wigner,PRB-WF}
\begin{equation}
\Delta\Phi = \frac{-e}{\varepsilon_0} \Delta P,
\end{equation}
where $e$ is the electron charge and $\varepsilon_0$ is the vacuum dielectric constant.
The surface dipole moment sets a barrier for the charge carriers. But
it does not change the position of the Fermi level of the electrode in the heterostrure.
Thus, the only change of the Fermi levels of the electrodes is caused by the aforementioned
charge transfer due to the Stark effect. 

Induced polarizations obtained for the top and bottom electrodes cause the change of
the work function by 1.53 eV for the anode and -1.50 eV for the cathode. 
The surface of the presented "square" elementary cell is smaller than that of the
molecular layer in the hexagonal lattice in Fig.~1 - by a factor of 0.9066 per
one molecule. This is due to the large "hole" within the molecular hexagons. 
Therefore, we estimate that the average change of the work function of the graphene
leads, applied to the layers proposed in Fig.~1, 
is around 1.39 eV and -1.36 eV for the anode and cathode, respectively.
Since the polarization induced in graphene near so much porous molecular lattice is homogeneous,
one can expect that the "effective" local change of the work function is much larger. 
The work function changes of the polarized graphene electrodes are not
so large when one compares them with these changes by about 2 eV in the two-dimensional
transition metal carbides and nitrides functionalized with OH \cite{OH}.

\subsection{Antiferroelectric arrangement, formation energies  
with graphene.}

\begin{figure}
\vspace{5mm}
\centerline{ \includegraphics[scale=0.19]{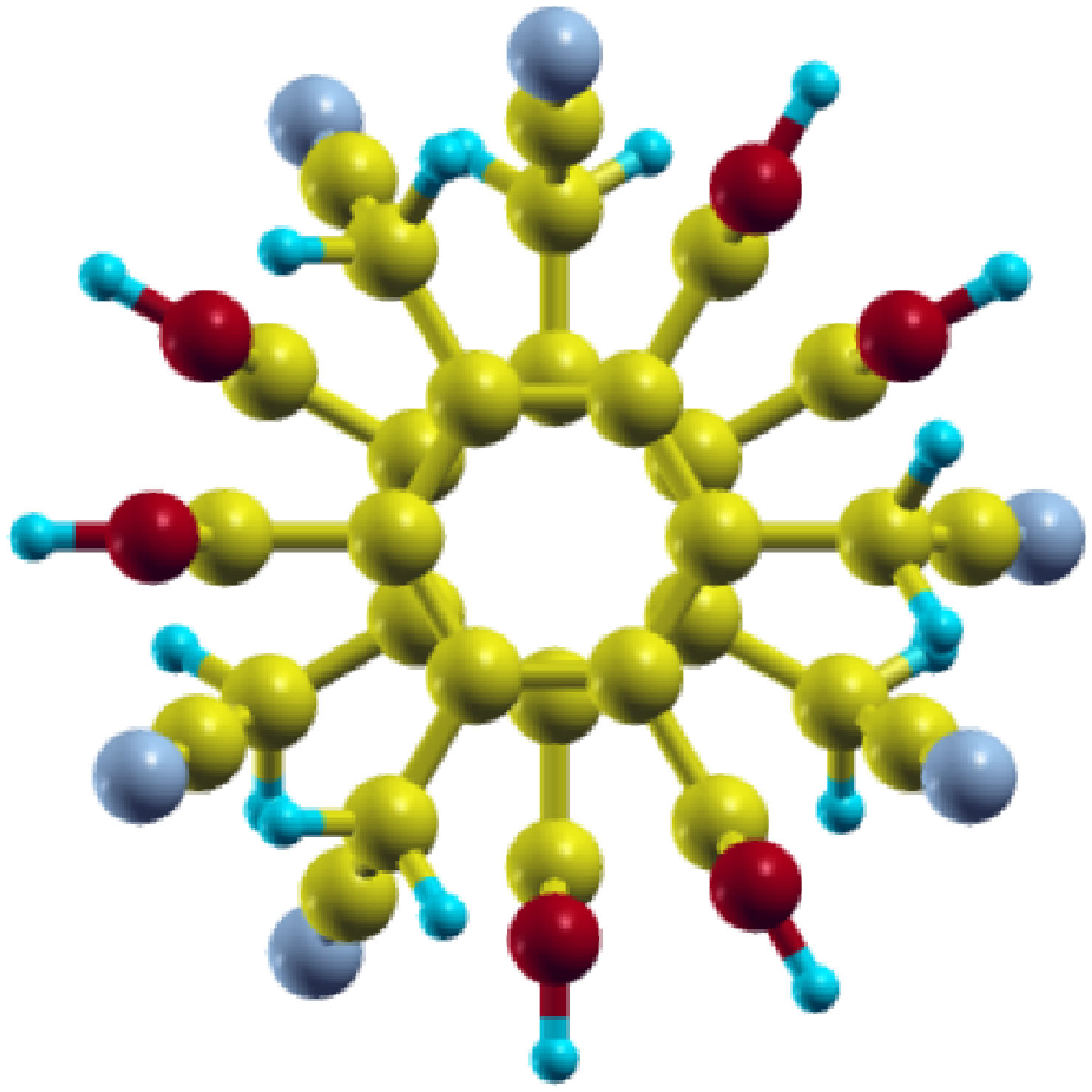}
             \includegraphics[scale=0.19]{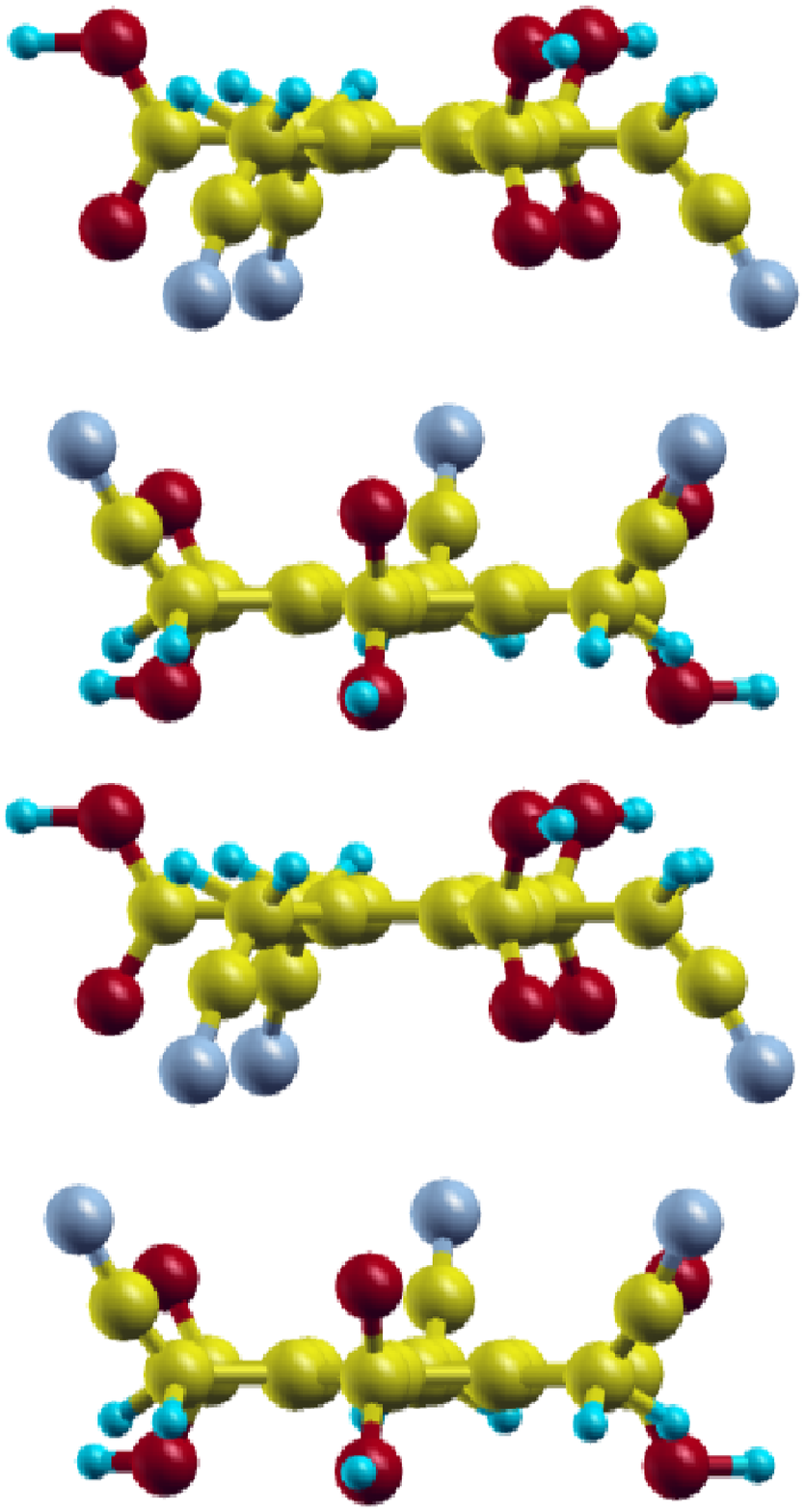}
             \includegraphics[scale=0.19]{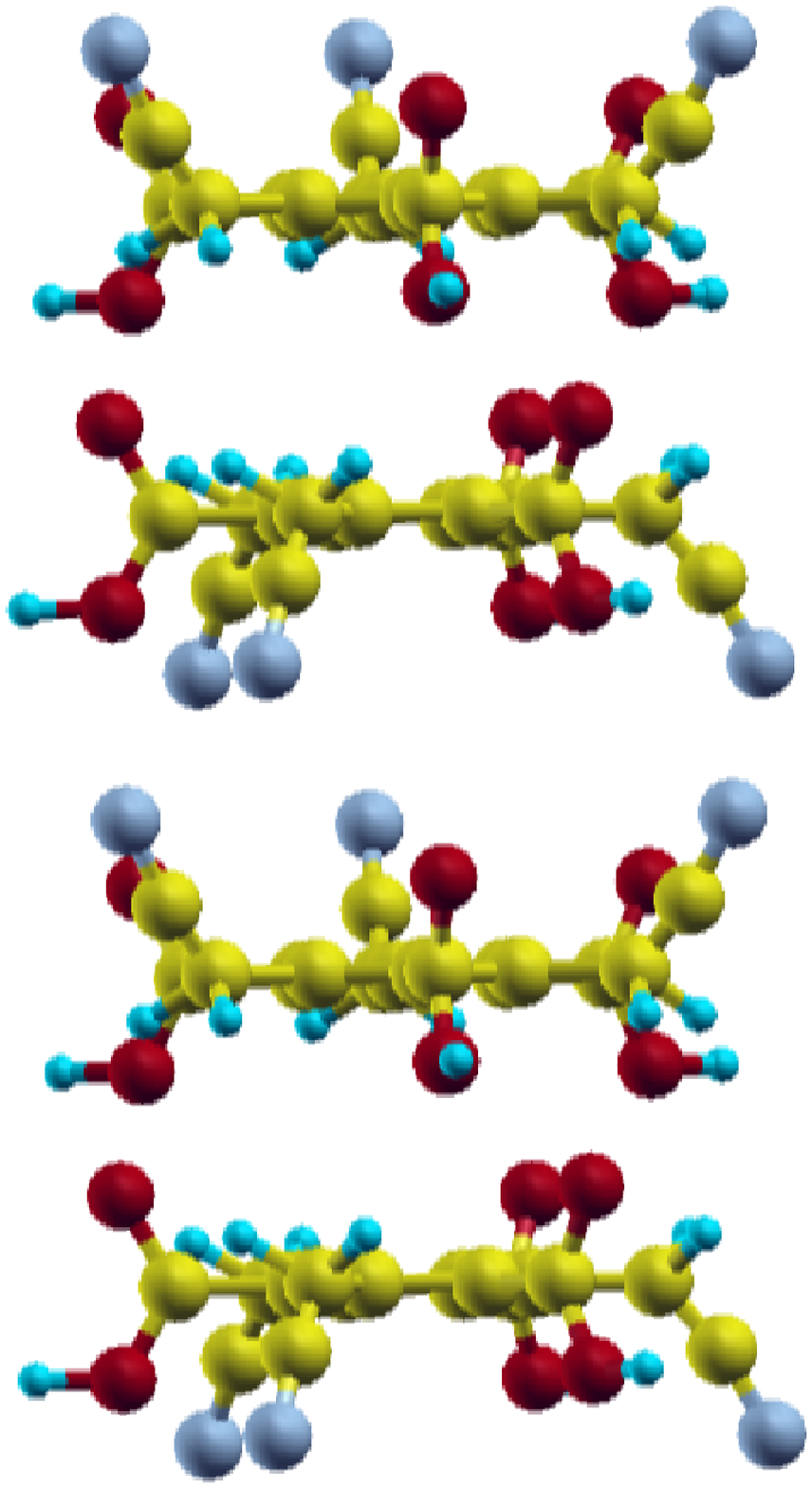}}\vspace{2mm}
\caption{The top (left panel) and side (middle and right panel) view 
on the two conformations of the antiferroelectric wire.} 
\label{S1}
\end{figure}

During the growth of the molecular crystal, it might happen that 
some rings adsorb at each other with the opposite side. In such case, locally,
the antiferroelectric phase will form. To estimate the probability of such event,
we calculated the 1D molecular wires oriented in: i) the ferroelectric, 
ii) the totally antiferroelectric (with total dipole moment 0) 
and iii) the not completely antiferroelectric way (with the total dipole moment lowered
with respect to the ferroelectric case but not equal to 0).
The two antiferroelectric cases are presented in Figure 8. They differ from each other
by a flip of H in the COOH groups. The orientation of the benzene centers in the ferroelectric
case (FE) is of the AA-stacking type (one on top of the other). In the antiferroelectric cases
(AFE), we rotated one molecule with respect to the other by 30 deg. This rotation does not
change the dipole moment along the molecular wire.
There is strong electrostatic repulsion within the planes rich of the H atoms
and the planes rich of the N and O atoms. For the first geometric conformation, 
the optimized distances between the central C-rings are 5.51 and 4.89 $\AA$,
for the N-rich and H-rich planes, respectively. The corresponding numbers for the
second geometric conformation are 5.55 and 4.85 $\AA$, respectively.
In both cases, the total energy per one molecule of the ferroelectric case is lower
than this energy of the antiferroelectric case: i) for the first conformation with vanishing
dipole moment, 
the FE-AFE energy difference is -675 meV and ii) for the the second conformation with a small
dipole moment,
the corresponding energy difference is -302 meV.    

The formation energy $E_f$, is defined as the difference of the total energies of the whole system
and the two subsystems, 
\begin{equation}
E_{f} = E_{A+B} - E_A -E_B. 
\end{equation}
For a single molecule on graphene,  
the formation energy is 1.33 eV for the bottom graphene layer (below the H-rich side of the molecule)
and 1.34 eV for the top graphene layer (above the N-rich side of the molecule). 
The formation of the ferroelectric wire is exotermic with 622 meV per molecule. While
for the totally antiferroelectric wire, it is endotermic with 53 meV per molecule.  
Therefore, the adsorption of the second and next layers on top of graphene
will be easier than that of the first layer. 

Detailed geometries of the molecule in vacuum, in the wire and on graphene are given
in Table 1. 

\begin{table}
\caption{Details of the geometry of the molecule in vacuum (M), molecule adsorbed 
with the H-side at graphene(M@g) and with the N-side (g@mol), as well as the
molecule in the ferroelectric (FE) and totally antiferroelectric wire (AFE).} 
\begin{tabular}{lccccc}
\hline \\[-0.2cm]
 bond lengths [$\AA$] & $\;\;$ M $\;\;$ & $\;\;$ M@g $\;\;$ & 
              $\;\;$ g@M $\;\;$  & $\;\;$  FE $\;\;$  & $\;\;$ AFE $\;$ \\[0.1cm]
 C-ring        & 1.405 & 1.434 & 1.434 & 1.405 & 1.412  \\
 C-ring-C(OOH) & 1.504 & 1.545 & 1.544 & 1.511 & 1.508  \\
 C-ring-C(CN)  & 1.527 & 1.525 & 1.524 & 1.523 & 1.528  \\
 C-C(N)        & 1.468 & 1.422 & 1.421 & 1.460 & 1.465  \\
 C-N           & 1.163 & 1.269 & 1.268 & 1.165 & 1.165   \\
 C-O(H)        & 1.370 & 1.389 & 1.390 & 1.358 & 1.376  \\
 C-O           & 1.215 & 1.171 & 1.170 & 1.218 & 1.209   \\
 O-H           & 0.984 & 1.004 & 1.004 & 0.986 & 0.987   \\
 C-H           & 1.102 & 1.151 & 1.150 & 1.100 & 1.104 \\[0.1cm]
 angles [deg] & M & M@g & g@M & FE & AFE \\[0.1cm]
 C-C-C(N)     & 116.7 & 113.2 & 113.2 & 114.1 & 118.0 \\
 C-C-N        & 175.9 & 166.1 & 166.2 & 178.4 & 173.3 \\
 O-C-O(H)     & 123.3 & 128.2 & 128.2 & 124.8 & 122.8  \\[0.1cm]
\hline
\end{tabular}
\label{occ}
\end{table}

\section{Conclusions}

We have shown that the organic ferroelectric layers with the lateral OH...O bonds
display a variety of desired properties for the photovoltaic efficiency. 
When these layers are the AA-type stacked, they are similar to
the columnar clusters with the dipole moment,  which were recently proposed as
the photovoltaic system without the p-n junction \cite{Sobol}.   
Ferroelectricity causes the cascade energy-level alignment of 
the hole and electron states of the subsequent molecular layers.
Importantly, the dipole moments of the layers from the middle of the stack do not
cancel, as it is in the bulk ferroelectrics where only the surface dipole moments resist. 
The fabrication of the presented layers could be done in the presence of the electric
field. Such that the ferroelectric (FE), and not the antiferroelectric (AFE), alignment of the
neighboring layers is achieved. Statistically, the mixed FE and AFE 
orderring occurs. However, the AFE orientation is energetically less favourable than the FE 
one by about 0.30-0.68 eV per molecule, depending on the partial or total vanishing of the
dipole moment originating from the COOH and CH2CN groups.    

Moreover, the donor or acceptor character of each layer depends on the direction from which
the charge carrier arrives, since this chemical property is defined with 
respect to the neighboring layers.
The optically generated hole-electron pair has the character of
the charge-transfer exciton, with an estimated Bohr radius of about 4.7 \AA.
The work function of the graphene electrodes, applied to such system, changes due to the
induced surface polarization - and it is about $\pm$1.4 eV for the anode and cathode leads, 
respectively, when a monolayer of molecules is sandwiched. Functionality of the presented   
system can be compared to that of the integrated heterojunctions \cite{Bein-2}.
The addition of the electron and hole transport layers is not necessary. 
Especially that these systems possess also a property of the separate paths 
for the electron and hole transport across the layers \cite{our-new}. \\

{\bf Acknowledgements} \\

This work has been supported by The National Science Centre of Poland
(the Projects No. 2013/11/B/ST3/04041 and DEC-2012/07/B/ST3/03412).
Calculations have been performed in the Cyfronet Computer Centre, using 
Prometheus computer which is a part of the PL-Grid Infrastrucure, 
and by part in the Interdisciplinary Centre of
Mathematical and Computer Modeling (ICM). \\ 
 \\

\end{document}